\documentclass[prl,twocolumn,floats,floatfix,aps,superscriptaddress,showpacs]{revtex4-1}

\usepackage{amsmath}
\usepackage{graphicx}
\usepackage{amsfonts}
\usepackage{amssymb}
\usepackage{color}
\usepackage{footnote}

\usepackage{bm}

\newcommand{\mC}{\mathcal{C}}

\newcommand{\be}{\begin{equation}}
\newcommand{\ee}{\end{equation}}

\newcommand{\bea}{\begin{eqnarray}}
\newcommand{\eea}{\end{eqnarray}}

\newcommand{\md}{\mu_{\textrm{d}}}
\newcommand{\ms}{\mu_{\textrm{s}}}
\newcommand{\fei}{f^{\textrm{ext}}_i}
\newcommand{\Ted}{T_{\textrm{Ed}}}
\newcommand{\Bed}{\beta_{\textrm{Ed}}}

\newcommand{\ve}{\varepsilon}

\DeclareMathOperator{\sgn}{sgn}

\begin{document}
 
\title{Generalized Edwards thermodynamics and marginal stability in a driven system with dry and viscous friction}

\author{Giacomo Gradenigo}
\affiliation{LIPHY, Universit\'e Grenoble Alpes and CNRS, F-38000 Grenoble, France}

\author{Eric Bertin}
\affiliation{LIPHY, Universit\'e Grenoble Alpes and CNRS, F-38000 Grenoble, France}

\date{\today}

\begin{abstract}
We consider a spring-block model with both dry and viscous frictions,
subjected to a periodic driving allowing mechanically stable configurations to be sampled.  We show that under strong driving, the scaling of the
correlation length with the energy density is incompatible with the
prediction of Edwards statistical approach, which assumes a uniform
sampling of mechanically stable configurations. A crossover between
the Edwards scaling and the non-standard high energy scaling is
observed at energy scales that depend on the viscous friction
coefficient.
Generalizing Edwards thermodynamics, we propose a statistical
framework, based on a sampling of marginally stable states,
that is able to describe the scaling of the correlation length in
the highly viscous regime.
\end{abstract}

\maketitle

The statistical description of driven dissipative systems remains one
of the challenging open issues of nonequilibrium statistical physics.
A subclass of these includes systems that are periodically driven and
relax to a mechanically stable configuration (MSC) between two driving
phases.  Of specific interest are such systems that, like granular
matter, are subject to dry friction, which generates a huge number of
MCSs, that can be characterized by an
extensive entropy.  Such systems are thus relevant candidates for
testing generalized forms of statistical mechanics.  In this spirit,
Edwards and coworkers~\cite{EO89,ME89,EM94,EG98,BKVS00,BHDC15} have put
forward the simplest generalization of equilibrium statistical
mechanics, by assuming that MSCs are
sampled uniformly (or according to an effective Boltzmann weight),
excluding configurations that are not
mechanically stable.  Whether this
simple assumption is valid or not has to be ultimately tested in
experiments or in numerical simulations, provided a driving protocol
is given.  Several tests of the Edwards hypothesis have been attempted
in packings of grains, both
experimentally~\cite{NKBJN98,SGS05,LCDB06,NRRCD09} and
numerically~\cite{KM02,M04,MD05,PCN06,BK15}. Tests have also been
performed in abstract models like spin and lattice gas
models~\cite{BPS00,LD01,L02,BFS02,DGL02,DGL03}, as well as in glass
and spin-glass models~\cite{CN00,BKVS00,BKVS01,DL01,LD03}.  Such tests
are performed by comparing the average values of some observables
recorded along the dynamics, with the values obtained from the flat
average over MSCs.  Note that, while the
original Edwards construction is based on volume and energy in analogy
to equilibrium statistical mechanics, another formulation focusing on
the stress tensor has also been put forward more recently
\cite{HHC07,HC09,BE09,BJE12,BZBC13,Daniels13}.
Overall, the Edwards assumption is generally believed to be a
reasonable approximation in most cases~\cite{BHDC15}, even though some
departure from the uniform sampling have been shown in some abstract
solvable models \cite{DGL02,DGL03}.  The complexity of Edwards
thermodynamics then mainly boils down to the computation of the
entropy (or free-energy) characterizing blocked states
\cite{BE03,BE06,BSWM08,WSJM11,APF14}.  A usual way to tackle this
difficult calculation is to resort either to simple abstract
models~\cite{BPS00,LD01,L02,BFS02,DGL02,DGL03}, or to
mean-field~\cite{SL03} or more involved~\cite{BE03} approximations.

Recently, however, a full treatment of the Edwards thermodynamics has
been performed in a more realistic spring-block model with dry
friction, showing the build-up of extended spatial correlations when
the strength of the driving is increased \cite{Gradenigo15}.
Here, we generalize the above model to include both viscous
and dry friction. The competition between viscous and dry friction has
been shown to play an essential role in the rheology of dense
suspensions \cite{Hebraud03,Mari13,Mari15},
and it is thus of high interest to try
to develop theoretical approaches able to take into account both
effects.  From a conceptual viewpoint, adding viscous friction is
actually a challenging test of Edwards thermodynamics: since viscous
friction affects only relaxation and not MSCs
(which are only controled by static dry friction), it appears as a key
ingredient controling the way MSCs are sampled.
Hence, any significant variation of
statistical properties as a function of the viscous damping
coefficient undoubtly shows that Edwards assumption fails to describe
in a faithful way the properties of the system.  Studying this
generalized spring-block model, we indeed find strong deviations from
the predictions of the standard Edwards approach.
The goal of this Letter is to present an extension of the Edwards theory
based on a non-uniform sampling of MSCs, emphasizing the
  importance of marginally stable states.
We show that this extended statistical framework is able to capture the main results of the numerical simulations of the spring-block model in the presence of viscous friction.

We consider a model represented by a one-dimensional chain of blocks of mass
$m$ connected by $N$ harmonic springs sliding on a horizontal
plane~\cite{BK67,CL89,GB10,BPG11,BGP14,Gradenigo15}.  Each particle is subjected
both to dry (Coulomb) friction and to viscous friction.  The position
of the $i^{\rm th}$-mass is denoted as $x_i$.  When sliding, a block
is subjected to a dissipative force proportional to the dynamic
friction coefficient, $f_{i,\textrm{dry}} = -\md mg \sgn(\dot{x}_i)$,
with $g$ the gravitational constant, and to a dissipative force
proportional to the viscous friction, $f_{i,\textrm{visc}} = -\gamma
\dot{x}_i$ (the dot denotes a time derivative).
When a block is at rest, it starts moving when the applied force
exceeds the static friction force, $|f_i|> \ms mg$.  The
elongation of the $i$-th spring is $\xi_i = x_i-x_{i-1}-l_0$, with
$l_0$ the constant rest length, so that the elastic force on each
block reads $ k (\xi_{i+1}-\xi_i)$, with $k$ the spring stiffness.
Taking $\sqrt{k/m}$, $gk/m$ and $mg$ as units of time, length and force
respectively, we can write the following dimensionless equation of motion: 
\be \label{eqofmotion}
\ddot{x}_i = - \gamma
\dot{x}_i -\md \sgn(\dot{x}_i) + x_{i+1}+x_{i-1}-2x_i + \fei ,
\ee
with $|\xi_{i+1}-\xi_i +\fei| > \ms$ the condition to start motion.
We simulated a chain of $N+1=4096$ blocks with open boundary
conditions taking an identical value of static and dynamic dry
friction coefficients, $\ms = \md$.
In the following, we do not distinguish between $\ms$ and $\md$,
and simply denote as $\mu$ the dry friction coefficient.

The ``blocked'' configurations are those which, in the absence of
external force, are mechanically stable: $\forall~i$, $\dot{x}_i=0$
and $|\xi_{i+1}-\xi_i|<\mu$.  We then define the following
\emph{tapping} dynamics: the external forces $\fei$ are switched on in
Eq.~(\ref{eqofmotion}) and act during a given period of time $\tau$,
after which they are switched off and the system relaxes to a MSC.
This procedure, that we call \emph{driving cycle}, is repeated a large
number of times to sample MSCs. The driving protocol 
consists in pulling a finite fraction of the particles, fixed to
$\rho=0.5$, with a constant force $F$, while keeping fixed the
duration $\tau$. Each MSC is characterized by the typical value of the
energy stored by the springs $\ve=(1/2N)\sum_{i=1}^N\xi_i^2$. \\

In the case where only dry friction is present, it has been shown that 
correlations of spring elongations, defined as $C(r)=\langle
\xi_{i+r}\xi_i \rangle/\langle \xi^2 \rangle$, extend over a correlation length which grows linearly with the energy density $\ve$ \cite{Gradenigo15}.
The Edwards approach is able to reproduce this scaling
of the correlation length with the energy density \cite{Gradenigo15}.
The Edwards ansatz for the probability of a configuration $\mC$ reads 
$P(\mC) = e^{-\Bed E(\mC)}\, \mathcal{F}(\mC)/\mathcal{Z}$, with 
$\Bed$ an effective temperature, $E(\mC)$
the energy of configuration $\mC$, and $\mathcal{Z}$ a normalization constant.
The function $\mathcal{F}(\mC)$ enforces the constraint of mechanical stability:
$\mathcal{F}(\mC)=1$ if $\mC$ is mechanically stable, and $\mathcal{F}(\mC)=0$
otherwise.
For the spring-block model, $\mathcal{F}(\mC) = \prod_{i=1}^{N-1}
\Theta(\mu-|\xi_{i+1}-\xi_i|)$ \cite{Gradenigo15}, with $\Theta$ the Heaviside function. 
By taking the continuum limit where the spring index $i$ is replaced by a continuous variable $s$ so that spring-elongations are represented as the local field $\xi(s)$, the
probability of a configuration reads as $e^{-S[{\bm \xi}]}$, with
(as a lowest order approximation)
a Gaussian effective Hamiltionian $ S[{\bm \xi}] = \int ds
[(\partial \xi /\partial s)^2/(4\mu^2) + \Bed\xi^2(s)/2]$ \cite{Gradenigo15}. 
Two important predictions of this theory are:
(i) The linear increase of the
correlation length $\lambda(\ve) \sim \ve$ with the average energy per
spring \cite{Gradenigo15};
(ii) The linear increase of the mean square displacement for the spring
elongation measured along the chain, $\langle[\Delta \xi(r)]^2\rangle
\sim r $, where $\Delta \xi(r) = \xi_{i+r}-\xi_i$.
These behaviors are modified in the presence of viscous friction.
\begin{figure}[t!]
  \includegraphics[width=\columnwidth]{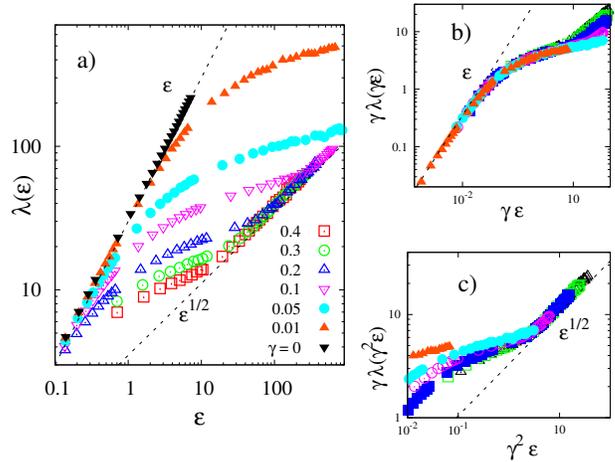}
\caption{a) Correlation length $\lambda$ as function of the average energy 
density $\ve$ of the sampled MSCs, for different values of the viscous
friction coefficient $\gamma$, indicated by different symbols.
Dashed lines emphasize the linear ($\lambda \sim \ve$) and square-root
($\lambda \sim \sqrt{\ve}$) behaviors reached for low and high energies respectively. b) First rescaling $\gamma \lambda = F_1(\gamma \ve)$ around the departure from the linear regime.
c) Second rescaling $\gamma \lambda = F_2(\gamma^2 \ve)$ around the
onset of the square-root regime.}
\label{fig-correl-length}
\end{figure}
Fig.~\ref{fig-correl-length}a) displays the correlation length as a function of energy for
different values of the viscous friction coefficient $\gamma$.
At relatively low energies,
all curves have a linear behavior as in the absence of viscous friction;
Note that the prefactor is independent of $\gamma$.
Increasing the energy, one observes a crossover, with
an intermediate regime which depends on $\gamma$, to a high-energy
scaling $\lambda(\ve) \sim \sqrt{\ve}$, with here again a prefactor which is independent of $\gamma$.
In between these asymptotic regimes, the correlation length $\lambda$ exhibits a strong dependence on the viscous friction coefficient $\gamma$.
The dependence on $\gamma$ can be rationalized according to two distinct scaling regimes. A first regime $\lambda(\ve) = \gamma^{-1} F_1(\gamma \ve)$
[Fig.~\ref{fig-correl-length}b)] describes the departure from the low-energy linear regime $\lambda(\ve) \sim \ve$.
A second regime $\lambda(\ve) = \gamma^{-1} F_2(\gamma^2 \ve)$
[Fig.~\ref{fig-correl-length}c)] describes the convergence to the asymptotic high energy scaling $\lambda(\ve) \sim \sqrt{\ve}$.
In other words, the linear regime $\lambda(\ve) \sim \ve$ is valid
for $\ve \ll \ve_1^* \sim \gamma^{-1}$, and the square-root regime
$\lambda(\ve) \sim \sqrt{\ve}$ is valid for $\ve \gg \ve_2^* \sim \gamma^{-2}$.
The fact that the prefactors of the scaling functions $F_{1,2}$ are 
$\gamma^{-1}$ in both cases indicate that the correlation length scales as 
$\lambda \sim \gamma^{-1}$ in the whole intermediate regime
$\ve_1^* < \ve < \ve_2^*$.

\begin{figure}[t!]
  \includegraphics[width=\columnwidth]{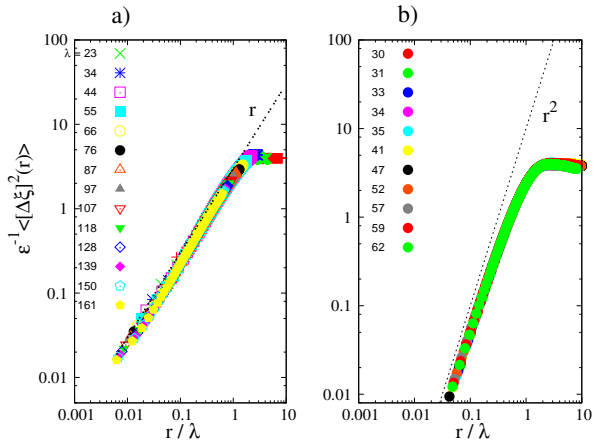}
  \caption{Mean square displacement of springs elongation,
  $\langle [\Delta \xi]^2(r) \rangle=\langle [ \xi_{i+r}-\xi_i]^2 \rangle $.
a) When only dry friction is present ($\gamma=0$), the mean-square displacement is diffusive, $\langle [\Delta \xi]^2(r) \rangle \sim r$.
b) For strong enough viscous friction, the mean-square displacement is ballistic, $\langle [\Delta \xi]^2(r) \rangle \sim r^2$ ($\gamma=0.3$).
Data are collapsed by plotting $\langle [\Delta \xi]^2(r) \rangle /\ve$
as a function of the rescaled distance $r/\lambda(\ve)$, with $\lambda(\ve)$ the correlation length.}
  \label{fig-MSD}
\end{figure}

Another characterization of the behavior of the model is through the mean-square displacement of the spring elongation $\langle[\Delta \xi(r)]^2\rangle$,
which was found to be linear (i.e., diffusive), $\langle[\Delta \xi(r)]^2\rangle \sim r$, when the dynamics involves only dry friction [Fig.~\ref{fig-MSD}a)].
In the presence of a strong enough viscous friction (or, for a given nonzero $\gamma$, at high enough energy), the mean-square displacement is observed to be ballistic,  $\langle[\Delta \xi(r)]^2\rangle \sim r^2$ [Fig.~\ref{fig-MSD}b)].

\begin{figure}[t!]
  \includegraphics[width=\columnwidth]{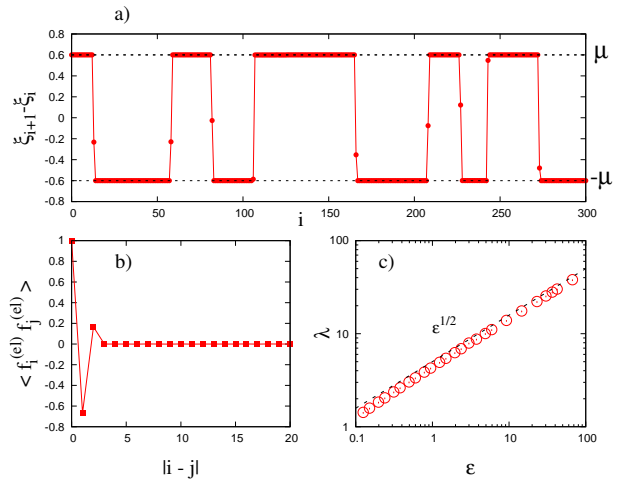}
  \caption{a) A typical MSC sampled at high energy in presence of viscous friction. The total force $f_i^{\rm el}=\xi_{i+1}-\xi_i$ acting on each mass is plotted as a function of the mass index ($\mu=0.6$).
b) Correlation of the elastic force $f_i$ at the end of the driving phase, before relaxation takes place.
c) Correlation length $\lambda$ of spring elongations, a function of the energy $\ve$, as obtained from the transfer operator method.
A Gaussian approximation (with standard deviation $\sigma=0.5$) of the delta function has been used.}
  \label{fig-config}
\end{figure}

The results obtained in the presence of viscous friction are clearly not compatible with those predicted in the standard Edwards framework, namely $\lambda(\ve) \sim \ve$ and $\langle[\Delta \xi(r)]^2\rangle \sim r$.
Let us emphasize that the presence of viscous friction only affects the relaxation process, and not the definition of MSCs, which depends only on dry friction.
The Edwards statistics is thus the same whatever the value of the viscous friction coefficient.
Hence the present results call for an alternative ansatz to describe the
non-uniform sampling of configurations in the presence of strong enough viscous damping. In order to determine such an ansatz, we start by
examining typical MSCs reached after a viscous relaxation, following a strong enough driving phase.
Fig.~\ref{fig-config}a) displays the total elastic force $f_i^{\rm el}=\xi_{i+1}-\xi_i$ acting on mass $i$ as a function of the mass index. Contrary to the dry friction case where the force spans in an essentially uniform way the interval $[-\mu,\mu]$ (in agreement with Edwards assumption), the force is seen to take almost everywhere only the two values $f_i^{\rm el}=\pm \mu$ [Fig.~\ref{fig-config}a)].
The typical length of the 'plateaus' at values $\pm \mu$ is of the order of the correlation length $\lambda$.

The emergence of such configurations of the force can be understood as follows in terms of the relaxation process.
At the end of the driving phase, the elastic forces $f_i^{\rm el}$ acting on different masses are uncorrelated [Fig~\ref{fig-config}b)]. Assuming a strong driving, the velocities are large in the initial stage of the relaxation, so that the dry friction term $-\mu \sgn(\dot{x}_i)$ can be neglected in this regime with respect to the viscous term
$-\gamma \dot{x}_i$. If $\gamma$ is large enough, we may also neglect inertia and use an overdamped dynamics.
In a continuum limit where the position $x_i(t)$ is replaced by a field
$x(s,t)$, where the continuous variable $s$ generalizes the index $i$, one obtains the following early-stage relaxational dynamics,
\be \label{eq:diffusive-relax}
\gamma \frac{\partial x}{\partial t} = \frac{\partial^2 x}{\partial s^2}.
\ee
This diffusive dynamics leads to a growth of the correlation length $\ell(t)$ of the field $x(s)$ as $\ell(t) \sim \sqrt{t}$.
This purely diffusive relaxation stops after a time $\sim t_{\rm rel}$, when velocities have decreased to a point where the viscous friction term becomes of the same order as the dry friction one.
For $t > t_{\rm rel}$, the dynamics reads
\be \label{eq:late-relax}
\gamma \frac{\partial x}{\partial t} = -\mu \sgn\left( \frac{\partial x}{\partial t} \right) + \frac{\partial^2 x}{\partial s^2}.
\ee
If the correlation length $\ell(t_{\rm rel})$ reached at the end of the diffusive relaxation is large enough, the intervals (in $s$) over which
$\partial x/\partial t$ has a constant sign remain large in the subsequent relaxation.
In a simplified picture, one may assume that these intervals do not change in time. Defining $\chi(s) = \sgn[\partial x/\partial t(s,t_{\rm rel})]$, one has
\be \label{eq:late-relax2}
\gamma \frac{\partial x}{\partial t} = -\mu \chi(s) + \frac{\partial^2 x}{\partial s^2}.
\ee
The relaxation described by Eq.~(\ref{eq:late-relax2}) converges to a MSC
$x^*(s)$ such that $d^2 x^*/ds^2=\mu \chi(s)$.
Since the elastic force $f^{\rm el}(s)$ acting on a mass with index $s$ is given
by $f^{\rm el}(s)=\partial^2 x/\partial s^2$, we end up with
$f^{\rm el}(s)=\mu \chi(s)$, thus recovering the typical shape of a configuration of the force shown on Fig.~\ref{fig-config}a).

Note that a piecewise constant force $f^{\rm el}(s)$ implies a piecewise linear elongation $\xi(s)$, since $d\xi/ds=f^{\rm el}(s)$. This piecewise linear behavior of the elongation in turn accounts for the ballistic behavior of the mean-square displacement $\langle[\Delta \xi(r)]^2\rangle \approx \mu^2 r^2$ of the elongation.
A simple scaling argument then allows one to understand in a simple way the origin of the behavior $\lambda \sim \sqrt{\ve}$ of the correlation length.
At large $r$, $\langle[\Delta \xi(r)]^2\rangle$ converges to
$2\langle \xi^2 \rangle = 4\ve$.
One thus expects $\langle[\Delta \xi(\lambda)]^2\rangle \approx 4\ve$,
which results, from the ballistic behavior, into $\lambda^2 \approx 4\ve/\mu^2$.
Note also that one recovers from this simple argument the fact that
$\lambda / \sqrt{\ve}$ is independent of $\gamma$ in this regime.

The fact that $\lambda \sim \gamma^{-1}$ in the intermediate scaling regime
[Fig.~\ref{fig-correl-length}] can be understood as follows. As argued above, the overdamped relaxation yields a correlation of the elastic force field. In contrast, an underdamped relaxation yields essentially no correlation of the elastic force, in agreement with the dry friction case.
The early stage of the relaxation is described by a linear equation, more conveniently expressed in Fourier space, introducing
$\hat{x}(q,t)=\int ds x(s,t)\, e^{iqs}$,
\be
\frac{\partial^2 \hat{x}}{\partial t^2}
+ \gamma \frac{\partial \hat{x}}{\partial t}
+ q^2 \hat{x}=0.
\ee
The solution of this equation takes the form, for $q \ll \gamma$,
\be \label{eq:sol}
\hat{x}(q,t) \approx X_1(q)\, e^{-tq^2/\gamma} + X_2(q)\, e^{-t (\gamma-q^2/\gamma)}
\ee
where $X_{1,2}(q)$ are related to the initial conditions.
When $\gamma$ is large (overdamped limit), the first term in the r.h.s.~of
Eq.~(\ref{eq:sol}) dominates the dynamics. For smaller values of $\gamma$, the second term comes into play, accounting for inertial effects.
The crossover between these two regimes is obtained by balancing the decay
rates, $q^2/\gamma \sim (\gamma-q^2/\gamma)$.
Taking $q \sim \lambda^{-1}$ as the relevant wavenumber, one obtains
that the crossover between inertial and overdamped regimes is reached for $\lambda \sim \gamma^{-1}$. This result is consistent with the numerical results reported in Fig.~\ref{fig-correl-length}, provided one identifies the inertial and overdamped regimes with the scaling regimes $\lambda \sim \ve$ and $\lambda \sim \sqrt{\ve}$ respectively.
Note that the existence of $\gamma$-independent regimes $\lambda \sim \ve$ and $\lambda \sim \sqrt{\ve}$ and of an intermediate regime where $\lambda \sim \gamma^{-1}$ is enough to account for the two scalings described by the functions
$F_{1,2}$ [Fig.~\ref{fig-correl-length}b) and c)].

To go beyond scaling arguments, we propose an ansatz generalizing the standard Edwards assumption of uniform sampling of MSCs.
Considering that MSCs typically sampled when viscous friction is high enough correspond to forces $f=\pm \mu$, we propose the following ansatz, which precisely enforces this property:
\be
P[{\boldsymbol \xi}] = \frac{1}{\mathcal{Z}} \, e^{- \frac{\Bed}{2} \sum_{i=1}^N \xi_i^2}
\prod_{i=1}^{N-1} \delta(\mu-|\xi_{i+1}-\xi_i|), 
\label{eq:new_Edwards}
\ee
where $\mathcal{Z}$ is a partition function determined by normalization,
\be \label{eq:partition}
\mathcal{Z} = \int d\xi_1 \dots d\xi_N \,
e^{- \frac{\Bed}{2} \sum_{i=1}^N \xi_i^2} \prod_{i=1}^{N-1} \delta(\mu-|\xi_{i+1}-\xi_i|), 
\ee
and where $\Bed=\Ted^{-1}$ is an effective inverse temperature.
Note that $\Bed$ is a parameter that can be eliminated at the end of the calculation, reexpressing all quantities in terms of the average energy density $\ve$.
In the following, we replace the delta functions in Eq.~(\ref{eq:partition})
by narrow Gaussian distributions of width $\sigma$.
Thermodynamic properties (free energy, average energy or entropy) as well as correlation functions can be determined
semi-analytically from Eqs.~(\ref{eq:new_Edwards}) and (\ref{eq:partition}), by evaluating the partition function $\mathcal{Z}$ using a transfer operator representation \cite{Gradenigo15},
$\mathcal{Z}=\textrm{Tr}(\mathcal{T}^N)$, where the linear operator
$\mathcal{T}$ acts on a function $\phi$ as
$\mathcal{T}[\phi](x)= \int dy\, T(x,y) \phi(y)$,
with $T(x,y)$ a symmetric $L^2$ kernel.
To evaluate $\mathcal{Z}$ as defined in Eq.~(\ref{eq:partition}), we use the kernel
\be
T(x,y)= e^{-\frac{\Bed}{4} (x^2+y^2)- [\mu^2-(x-y)^2]^2/(2\sigma^2)}.
\ee
Note that we have used here periodic boundary conditions, which does not affect the results in the thermodynamic limit.
The properties of the kernel $T(x,y)$ guarantees the existence of an
orthonormal set of eigenvectors of $\mathcal{T}$, which can be
numerically diagonalized. Following this approach we have checked that
our results do not depend on the value of
the parameter $\sigma$ in the large $\Ted$ limit.
The two-point correlation function $C(r)=\langle
\xi_{i+r}\xi_i \rangle/\langle \xi^2 \rangle$ can be numerically determined
within the transfer operator formalism from the eigenvectors of
$\mathcal{T}$, and from it the correlation length $\lambda(\ve)$ is
obtained (technical details on the transfer operator method can be found in the Supplemental Material of \cite{Gradenigo15}).
Extracting the correlation length from $C(r)$ for different values of the energy $\ve$, we recover the behavior $\lambda \sim \sqrt{\ve}$ [Fig.~\ref{fig-config}c)].
Note that the prefactor is independent of $\gamma$, since $\gamma$ does not appear in Eq.~(\ref{eq:new_Edwards}).

The above results suggest to consider, beyond the present specific model, the following prescription for systems subjected to both dry and viscous frictions.
Mechanical stability, as resulting from dry friction, is expressed by inequalities involving the dry friction coefficient. We call marginally stable the configurations such that these inequalities are satisfied as equalities.
A general formulation of the ansatz (\ref{eq:new_Edwards}) is that \emph{marginally stable configurations are sampled with a Boltzmann weight, while other configurations have zero probability}.

In summary, we have shown by studying a periodically driven spring-block model that the presence of viscous friction deeply changes the way MSCs are sampled,
yielding a scaling of the correlation length with energy density which is
incompatible with the Edwards assumption.
We have shown that typically sampled MSCs correspond to states with marginal mechanical stability, which provides another example of system where marginal stability plays a key role, in addition to the known examples of glasses and soft amorphous solids \cite{Wyart15}, notably in connection to the Gardner transition \cite{Zamponi14}.
We have proposed a generalized ansatz according to which only marginally stable MSCs have a non-zero probability, and are sampled according to an effective Boltzmann weight.
This ansatz is able to reproduce the key features of the spring-block model under viscous friction, including the square-root scaling of the correlation length with energy, and the ballistic behavior of the mean-square displacement of spring elongation.
It would be of interest to test this ansatz in other types of systems where viscous damping is present, like sedimenting suspensions under tapping dynamics.

\begin{acknowledgments}
The authors are grateful to J.-L. Barrat for many fruitful discussions.
G.G.~acknowledges Financial support from ERC Grant No. ADG20110209.
\end{acknowledgments}

\end{document}